\documentclass[a4paper,conference]{IEEEtran}
\IEEEoverridecommandlockouts
\usepackage{cite}
\usepackage{amsmath,amssymb,amsfonts}
\usepackage{algorithmic}
\usepackage{graphicx}
\usepackage{textcomp}
\usepackage{xcolor}
\usepackage{color,soul}
\usepackage{float}
\def\BibTeX{{\rm B\kern-.05em{\sc i\kern-.025em b}\kern-.08em
    T\kern-.1667em\lower.7ex\hbox{E}\kern-.125emX}}
\usepackage{graphicx}
\usepackage{subcaption}
\usepackage{booktabs}
\usepackage{placeins}
\usepackage{pgfplots}
\usepackage{geometry}
\usepackage[acronyms,nonumberlist,nopostdot,nomain,nogroupskip,acronymlists={hidden}]{glossaries}
\newglossary[algh]{hidden}{acrh}{acnh}{Hidden Acronyms}
\glsdisablehyper
\newacronym{3gpp}{3GPP}{3rd Generation Partnership Project}
\newacronym{4g}{4G}{4th generation mobile network}
\newacronym{5g}{5G}{5th generation mobile network}
\newacronym{6g}{6G}{6th generation mobile network}
\newacronym{nextg}{NextG}{Next Generation}
\newacronym{5gc}{5GC}{5G Core}
\newacronym{adc}{ADC}{Analog to Digital Converter}
\newacronym{aerpaw}{AERPAW}{Aerial Experimentation and Research Platform for Advanced Wireless}
\newacronym{ai}{AI}{Artificial Intelligence}
\newacronym{aimd}{AIMD}{Additive Increase Multiplicative Decrease}
\newacronym{am}{AM}{Acknowledged Mode}
\newacronym{amc}{AMC}{Adaptive Modulation and Coding}
\newacronym{amf}{AMF}{Access and Mobility Management Function}
\newacronym{aops}{AOPS}{Adaptive Order Prediction Scheduling}
\newacronym{api}{API}{Application Programming Interface}
\newacronym{apn}{APN}{Access Point Name}
\newacronym{aqm}{AQM}{Active Queue Management}
\newacronym{ausf}{AUSF}{Authentication Server Function}
\newacronym{avc}{AVC}{Advanced Video Coding}
\newacronym{awgn}{AGWN}{Additive White Gaussian Noise}
\newacronym{balia}{BALIA}{Balanced Link Adaptation Algorithm}
\newacronym{bbu}{BBU}{Base Band Unit}
\newacronym{bdp}{BDP}{Bandwidth-Delay Product}
\newacronym{ber}{BER}{Bit Error Rate}
\newacronym{bf}{BF}{Beamforming}
\newacronym{bler}{BLER}{Block Error Rate}
\newacronym{brr}{BRR}{Bayesian Ridge Regressor}
\newacronym{bsr}{BSR}{Buffer Status Report}
\newacronym{bs}{BS}{Base Station}
\newacronym{bpsk}{BPSK}{Binary Phase-shift keying}
\newacronym{bss}{BSS}{Business Support System}
\newacronym{ca}{CA}{Carrier Aggregation}
\newacronym{caas}{CaaS}{Connectivity-as-a-Service}
\newacronym{cb}{CB}{Code Block}
\newacronym{cc}{CC}{Congestion Control}
\newacronym{ccid}{CCID}{Congestion Control ID}
\newacronym{cco}{CC}{Carrier Component}
\newacronym{cd}{CD}{Continuous Delivery}
\newacronym{cdd}{CDD}{Cyclic Delay Diversity}
\newacronym{cdf}{CDF}{Cumulative Distribution Function}
\newacronym{cdma}{CDMA}{Code-Division Multiple Access}
\newacronym{cdn}{CDN}{Content Distribution Network}
\newacronym{ci}{CI}{Continuous Integration}
\newacronym{cicd}{CI/CD}{Continuous Integration/Continuous Delivery}
\newacronym{cir}{CIR}{Channel Impulse Response}
\newacronym{cn}{CN}{Core Network}
\newacronym{codel}{CoDel}{Controlled Delay Management}
\newacronym{comac}{COMAC}{Converged Multi-Access and Core}
\newacronym{cord}{CORD}{Central Office Re-architected as a Datacenter}
\newacronym{cornet}{CORNET}{COgnitive Radio NETwork}
\newacronym{cosmos}{COSMOS}{Cloud Enhanced Open Software Defined Mobile Wireless Testbed for City-Scale Deployment}
\newacronym{cots}{COTS}{Commercial Off-the-Shelf}
\newacronym{cp}{CP}{Control Plane}
\newacronym{cpu}{CPU}{Central Processing Unit}
\newacronym{cqi}{CQI}{Channel Quality Information}
\newacronym{cr}{CR}{Cognitive Radio}
\newacronym{cran}{CRAN}{Cloud \gls{ran}}
\newacronym{crs}{CRS}{Cell Reference Signal}
\newacronym{csi}{CSI}{Channel State Information}
\newacronym{csirs}{CSI-RS}{Channel State Information - Reference Signal}
\newacronym{cu}{CU}{Central Unit}
\newacronym{d2tcp}{D$^2$TCP}{Deadline-aware Data center TCP}
\newacronym{d3}{D$^3$}{Deadline-Driven Delivery}
\newacronym{dac}{DAC}{Digital to Analog Converter}
\newacronym{dag}{DAG}{Directed Acyclic Graph}
\newacronym{darpa}{DARPA}{Defense Advanced Research Projects Agency}
\newacronym{das}{DAS}{Distributed Antenna System}
\newacronym{dash}{DASH}{Dynamic Adaptive Streaming over HTTP}
\newacronym{dc}{DC}{Dual Connectivity}
\newacronym{dccp}{DCCP}{Datagram Congestion Control Protocol}
\newacronym{dce}{DCE}{Direct Code Execution}
\newacronym{dci}{DCI}{Downlink Control Information}
\newacronym{dcl}{DCL}{Dear Colleague Letter}
\newacronym{dctcp}{DCTCP}{Data Center TCP}
\newacronym{devops}{DevOps}{Development and Operations}
\newacronym{dl}{DL}{Deep Learning}
\newacronym{dmr}{DMR}{Deadline Miss Ratio}
\newacronym{dmrs}{DMRS}{DeModulation Reference Signal}
\newacronym{drlcc}{DRL-CC}{Deep Reinforcement Learning Congestion Control}
\newacronym{drs}{DRS}{Discovery Reference Signal}
\newacronym{dt}{DT}{Digital Twin}
\newacronym{dtn}{DTN}{Digital Twin Network}
\newacronym{dtmn}{DTMN}{Digital Twin for Mobile Network}
\newacronym{dtwn}{DTWN}{Digital Twin Wireless Network}
\newacronym{du}{DU}{Distributed Unit}
\newacronym{e2e}{E2E}{end-to-end}
\newacronym{ecaas}{ECaaS}{Edge-Cloud-as-a-Service}
\newacronym{ecn}{ECN}{Explicit Congestion Notification}
\newacronym{edf}{EDF}{Earliest Deadline First}
\newacronym{em}{EM}{Electro-Magnetic}
\newacronym{embb}{eMBB}{Enhanced Mobile Broadband}
\newacronym{empower}{EMPOWER}{EMpowering transatlantic PlatfOrms for advanced WirEless Research}
\newacronym{enb}{eNB}{evolved Node Base}
\newacronym{endc}{EN-DC}{E-UTRAN-\gls{nr} \gls{dc}}
\newacronym{epc}{EPC}{Evolved Packet Core}
\newacronym{eps}{EPS}{Evolved Packet System}
\newacronym{es}{ES}{Edge Server}
\newacronym{etsi}{ETSI}{European Telecommunications Standards Institute}
\newacronym[firstplural=Estimated Times of Arrival (ETAs)]{eta}{ETA}{Estimated Time of Arrival}
\newacronym{eutran}{E-UTRAN}{Evolved Universal Terrestrial Access Network}
\newacronym{faas}{FaaS}{Function-as-a-Service}
\newacronym{fapi}{FAPI}{Functional Application Platform Interface}
\newacronym{fcc}{FCC}{Federal Communications Commission}
\newacronym{fdd}{FDD}{Frequency Division Duplexing}
\newacronym{fdm}{FDM}{Frequency Division Multiplexing}
\newacronym{fdma}{FDMA}{Frequency Division Multiple Access}
\newacronym{fed4fire}{FED4FIRE+}{Federation 4 Future Internet Research and Experimentation Plus}
\newacronym{fir}{FIR}{Finite Impulse Response}
\newacronym{fit}{FIT}{Future \acrlong{iot}}
\newacronym{fl}{FL}{Federated Learning}
\newacronym{fpga}{FPGA}{Field Programmable Gate Array}
\newacronym{fr2}{FR2}{Frequency Range 2}
\newacronym{fs}{FS}{Fast Switching}
\newacronym{fscc}{FSCC}{Flow Sharing Congestion Control}
\newacronym{ftp}{FTP}{File Transfer Protocol}
\newacronym{fw}{FW}{Flow Window}
\newacronym{ga128}{Ga}{Golay Sequence type A}
\newacronym{ge}{GE}{Gaussian Elimination}
\newacronym{glfsr}{GLFSR}{Galois Linear Feedback Shift Register}
\newacronym{gnb}{gNB}{Next Generation Node Base}
\newacronym{gold}{Gold}{Gold}
\newacronym{gop}{GOP}{Group of Pictures}
\newacronym{gpr}{GPR}{Gaussian Process Regressor}
\newacronym{gpu}{GPU}{Graphics Processing Unit}
\newacronym{gtp}{GTP}{GPRS Tunneling Protocol}
\newacronym{gtpc}{GTP-C}{GPRS Tunnelling Protocol Control Plane}
\newacronym{gtpu}{GTP-U}{GPRS Tunnelling Protocol User Plane}
\newacronym{gtpv2c}{GTPv2-C}{\gls{gtp} v2 - Control}
\newacronym{gw}{GW}{Gateway}
\newacronym{harq}{HARQ}{Hybrid Automatic Repeat reQuest}
\newacronym{hetnet}{HetNet}{Heterogeneous Network}
\newacronym{hh}{HH}{Hard Handover}
\newacronym{hol}{HOL}{Head-of-Line}
\newacronym{hqf}{HQF}{Highest-quality-first}
\newacronym{hss}{HSS}{Home Subscription Server}
\newacronym{http}{HTTP}{HyperText Transfer Protocol}
\newacronym{ia}{IA}{Initial Access}
\newacronym{iab}{IAB}{Integrated Access and Backhaul}
\newacronym{ic}{IC}{Incident Command}
\newacronym{ietf}{IETF}{Internet Engineering Task Force}
\newacronym{ifw}{IFW}{Interference Free Window}
\newacronym{imsi}{IMSI}{International Mobile Subscriber Identity}
\newacronym{imt}{IMT}{International Mobile Telecommunication}
\newacronym{iot}{IoT}{Internet of Things}
\newacronym{ip}{IP}{Internet Protocol}
\newacronym{iq}{IQ}{In-phase and Quadrature}
\newacronym{isi}{ISI}{Inter-Symbol Interference}
\newacronym{itu}{ITU}{International Telecommunication Union}
\newacronym{kpi}{KPI}{Key Performance Indicator}
\newacronym{kvm}{KVM}{Kernel-based Virtual Machine}
\newacronym{lfsr}{LFSR}{Linear Feedback Shift Register}
\newacronym{los}{LOS}{Line-of-Sight}
\newacronym{ls}{LS}{Loosely Synchronised}
\newacronym{lsm}{LSM}{Link-to-System Mapping}
\newacronym{lstm}{LSTM}{Long Short Term Memory}
\newacronym{lte}{LTE}{Long Term Evolution}
\newacronym{lxc}{LXC}{Linux Container}
\newacronym{m2m}{M2M}{Machine to Machine}
\newacronym{mac}{MAC}{Medium Access Control}
\newacronym{mai}{MAI}{Multiple Access Interference}
\newacronym{manet}{MANET}{Mobile Ad Hoc Network}
\newacronym{mano}{MANO}{Management and Orchestration}
\newacronym{mc}{MC}{Multi-Connectivity}
\newacronym{mcc}{MCC}{Mobile Cloud Computing}
\newacronym{mchem}{MCHEM}{Massive Channel Emulator}
\newacronym{mcs}{MCS}{Modulation and Coding Scheme}
\newacronym{mec}{MEC}{Multi-access Edge Computing}
\newacronym{mec2}{MEC}{Mobile Edge Cloud}
\newacronym{mec3}{MEC}{Mobile Edge Computing}
\newacronym{mfc}{MFC}{Mobile Fog Computing}
\newacronym{mi}{MI}{Mutual Information}
\newacronym{mib}{MIB}{Master Information Block}
\newacronym{miesm}{MIESM}{Mutual Information Based Effective SINR}
\newacronym{mimo}{MIMO}{Multiple Input, Multiple Output}
\newacronym{mgen}{MGEN}{Multi-Generator}
\newacronym{ml}{ML}{Machine Learning}
\newacronym{mlr}{MLR}{Maximum-local-rate}
\newacronym[plural=\gls{mme}s,firstplural=Mobility Management Entities (MMEs)]{mme}{MME}{Mobility Management Entity}
\newacronym{mmtc}{mMTC}{Massive Machine-Type Communications}
\newacronym{mmwave}{mmWave}{millimeter wave}
\newacronym{mpdccp}{MP-DCCP}{Multipath Datagram Congestion Control Protocol}
\newacronym{mptcp}{MPTCP}{Multipath TCP}
\newacronym{mr}{MR}{Maximum Rate}
\newacronym{mrdc}{MR-DC}{Multi \gls{rat} \gls{dc}}
\newacronym{mse}{MSE}{Mean Square Error}
\newacronym{mss}{MSS}{Maximum Segment Size}
\newacronym{mt}{MT}{Mobile Termination}
\newacronym{mtd}{MTD}{Machine-Type Device}
\newacronym{mtu}{MTU}{Maximum Transmission Unit}
\newacronym{mumimo}{MU-MIMO}{Multi-user \gls{mimo}}
\newacronym{mvno}{MVNO}{Mobile Virtual Network Operator}
\newacronym{nalu}{NALU}{Network Abstraction Layer Unit}
\newacronym{nas}{NAS}{Network Attached Storage}
\newacronym{nbiot}{NB-IoT}{Narrow Band IoT}
\newacronym{nfv}{NFV}{Network Function Virtualization}
\newacronym{nfvi}{NFVI}{Network Function Virtualization Infrastructure}
\newacronym{nic}{NIC}{Network Interface Card}
\newacronym{nlos}{NLOS}{Non-Line-of-Sight}
\newacronym{now}{NOW}{Non Overlapping Window}
\newacronym{nrdz}{NRDZ}{National Radio Dynamic Zone}
\newacronym{nsf}{NSF}{National Science Foundation}
\newacronym{nsm}{NSM}{Network Service Mesh}
\newacronym[type=hidden]{nr}{NR}{New Radio}
\newacronym{nrf}{NRF}{Network Repository Function}
\newacronym{nsa}{NSA}{Non Stand Alone}
\newacronym{nse}{NSE}{Network Slicing Engine}
\newacronym{nssf}{NSSF}{Network Slice Selection Function}
\newacronym{ntp}{NTP}{Network Time Protocol}
\newacronym{o2i}{O2I}{Outdoor to Indoor}
\newacronym{oai}{OAI}{OpenAirInterface}
\newacronym{oaicn}{OAI-CN}{\gls{oai} \acrlong{cn}}
\newacronym{oairan}{OAI-RAN}{\acrlong{oai} \acrlong{ran}}
\newacronym{oam}{OAM}{Operations, Administration and Maintenance}
\newacronym[plural=\gls{obu}s,firstplural=Onboard Units (OBUs)]{obu}{OBU}{Onboard Unit}
\newacronym{ofdm}{OFDM}{Orthogonal Frequency Division Multiplexing}
\newacronym{olia}{OLIA}{Opportunistic Linked Increase Algorithm}
\newacronym{omec}{OMEC}{Open Mobile Evolved Core}
\newacronym{onap}{ONAP}{Open Network Automation Platform}
\newacronym{onf}{ONF}{Open Networking Foundation}
\newacronym{onos}{ONOS}{Open Networking Operating System}
\newacronym{oom}{OOM}{\gls{onap} Operations Manager}
\newacronym{opnfv}{OPNFV}{Open Platform for \gls{nfv}}
\newacronym[type=hidden]{oran}{O-RAN}{Open \gls{ran}}
\newacronym{orbit}{ORBIT}{Open-Access Research Testbed for Next-Generation Wireless Networks}
\newacronym{os}{OS}{Operating System}
\newacronym{osm}{OSM}{Open Street Map}
\newacronym{oss}{OSS}{Operations Support System}
\newacronym{pa}{PA}{Position-aware}
\newacronym{pase}{PASE}{Prioritization, Arbitration, and Self-adjusting Endpoints}
\newacronym{pawr}{PAWR}{Platforms for Advanced Wireless Research}
\newacronym{pbch}{PBCH}{Physical Broadcast Channel}
\newacronym{pcef}{PCEF}{Policy and Charging Enforcement Function}
\newacronym{pcfich}{PCFICH}{Physical Control Format Indicator Channel}
\newacronym{pcrf}{PCRF}{Policy and Charging Rules Function}
\newacronym{pdcch}{PDCCH}{Physical Downlink Control Channel}
\newacronym{pdcp}{PDCP}{Packet Data Convergence Protocol}
\newacronym{pdsch}{PDSCH}{Physical Downlink Shared Channel}
\newacronym{pdu}{PDU}{Packet Data Unit}
\newacronym{pdp}{PDP}{Power Delay Profile}
\newacronym{pf}{PF}{Proportional Fair}
\newacronym{pgw}{PGW}{Packet Gateway}
\newacronym{phich}{PHICH}{Physical Hybrid ARQ Indicator Channel}
\newacronym{phy}{PHY}{Physical}
\newacronym{pl}{PL}{Path Loss}
\newacronym{pmch}{PMCH}{Physical Multicast Channel}
\newacronym{pmi}{PMI}{Precoding Matrix Indicators}
\newacronym{powder}{POWDER}{Platform for Open Wireless Data-driven Experimental Research}
\newacronym{ppo}{PPO}{Proximal Policy Optimization}
\newacronym{ppp}{PPP}{Poisson Point Process}
\newacronym{prach}{PRACH}{Physical Random Access Channel}
\newacronym{prb}{PRB}{Physical Resource Block}
\newacronym{psnr}{PSNR}{Peak Signal to Noise Ratio}
\newacronym{pss}{PSS}{Primary Synchronization Signal}
\newacronym{pucch}{PUCCH}{Physical Uplink Control Channel}
\newacronym{pusch}{PUSCH}{Physical Uplink Shared Channel}
\newacronym{qam}{QAM}{Quadrature Amplitude Modulation}
\newacronym{qci}{QCI}{\gls{qos} Class Identifier}
\newacronym{qoe}{QoE}{Quality of Experience}
\newacronym{qos}{QoS}{Quality of Service}
\newacronym{qtgui}{QT-GUI}{QT Graphical User Interface}
\newacronym{quic}{QUIC}{Quick UDP Internet Connections}
\newacronym{rach}{RACH}{Random Access Channel}
\newacronym{ran}{RAN}{Radio Access Network}
\newacronym[firstplural=Radio Access Technologies (RATs)]{rat}{RAT}{Radio Access Technology}
\newacronym{rcn}{RCN}{Research Coordination Network}
\newacronym{rec}{REC}{Radio Edge Cloud}
\newacronym{red}{RED}{Random Early Detection}
\newacronym{renew}{RENEW}{Reconfigurable Eco-system for Next-generation End-to-end Wireless}
\newacronym{rf}{RF}{Radio Frequency}
\newacronym{rfc}{RFC}{Request for Comments}
\newacronym{rfr}{RFR}{Random Forest Regressor}
\newacronym{ric}{RIC}{\gls{ran} Intelligent Controller}
\newacronym{rlc}{RLC}{Radio Link Control}
\newacronym{rlf}{RLF}{Radio Link Failure}
\newacronym{rlnc}{RLNC}{Random Linear Network Coding}
\newacronym{rmse}{RMSE}{Root Mean Squared Error}
\newacronym{rnis}{RNIS}{Radio Network Information Service}
\newacronym{rr}{RR}{Round Robin}
\newacronym{rrc}{RRC}{Radio Resource Control}
\newacronym{rrm}{RRM}{Radio Resource Management}
\newacronym{rru}{RRU}{Remote Radio Unit}
\newacronym{rs}{RS}{Remote Server}
\newacronym{rsrp}{RSRP}{Reference Signal Received Power}
\newacronym{rsrq}{RSRQ}{Reference Signal Received Quality}
\newacronym{rss}{RSS}{Received Signal Strength}
\newacronym{rssi}{RSSI}{Received Signal Strength Indicator}
\newacronym{rsu}{RSU}{Road-Side Unit}
\newacronym{rtt}{RTT}{Round Trip Time}
\newacronym{ru}{RU}{Radio Unit}
\newacronym{rw}{RW}{Receive Window}
\newacronym{rx}{RX}{Receiver}
\newacronym{s1ap}{S1AP}{S1 Application Protocol}
\newacronym{sa}{SA}{standalone}
\newacronym{sack}{SACK}{Selective Acknowledgment}
\newacronym{sap}{SAP}{Service Access Point}
\newacronym{sc2}{SC2}{Spectrum Collaboration Challenge}
\newacronym{scef}{SCEF}{Service Capability Exposure Function}
\newacronym{sch}{SCH}{Secondary Cell Handover}
\newacronym{scoot}{SCOOT}{Split Cycle Offset Optimization Technique}
\newacronym{sctp}{SCTP}{Stream Control Transmission Protocol}
\newacronym{sdap}{SDAP}{Service Data Adaptation Protocol}
\newacronym{sd}{SD}{Standard Deviation}
\newacronym{sdk}{SDK}{Software Development Kit}
\newacronym{sdm}{SDM}{Space Division Multiplexing}
\newacronym{sdma}{SDMA}{Spatial Division Multiple Access}
\newacronym{sdn}{SDN}{Software-defined Networking}
\newacronym{sdr}{SDR}{Software-defined Radio}
\newacronym{seba}{SEBA}{SDN-Enabled Broadband Access}
\newacronym{sgsn}{SGSN}{Serving GPRS Support Node}
\newacronym{sgw}{SGW}{Service Gateway}
\newacronym{si}{SI}{Study Item}
\newacronym{sib}{SIB}{Secondary Information Block}
\newacronym{sinr}{SINR}{Signal to Interference plus Noise Ratio}
\newacronym{sip}{SIP}{Session Initiation Protocol}
\newacronym{siso}{SISO}{Single Input, Single Output}
\newacronym{sla}{SLA}{Service Level Agreement}
\newacronym{sm}{SM}{Saturation Mode}
\newacronym{smf}{SMF}{Session Management Function}
\newacronym{smo}{SMO}{Service Management and Orchestration}
\newacronym{sms}{SMS}{Short Message Service}
\newacronym{smsgmsc}{SMS-GMSC}{\gls{sms}-Gateway}
\newacronym{snr}{SNR}{Signal-to-Noise-Ratio}
\newacronym{son}{SON}{Self-Organizing Network}
\newacronym{sptcp}{SPTCP}{Single Path TCP}
\newacronym{srb}{SRB}{Service Radio Bearer}
\newacronym{srn}{SRN}{Standard Radio Node}
\newacronym{srs}{SRS}{Sounding Reference Signal}
\newacronym{ss}{SS}{Synchronization Signal}
\newacronym{sss}{SSS}{Secondary Synchronization Signal}
\newacronym{st}{ST}{Spanning Tree}
\newacronym{svc}{SVC}{Scalable Video Coding}
\newacronym{tb}{TB}{Transport Block}
\newacronym{tcp}{TCP}{Transmission Control Protocol}
\newacronym{tdd}{TDD}{Time Division Duplexing}
\newacronym{tdm}{TDM}{Time Division Multiplexing}
\newacronym{tdma}{TDMA}{Time Division Multiple Access}
\newacronym{tfl}{TfL}{Transport for London}
\newacronym{tfrc}{TFRC}{TCP-Friendly Rate Control}
\newacronym{tft}{TFT}{Traffic Flow Template}
\newacronym{tgen}{TGEN}{Traffic Generator}
\newacronym{tip}{TIP}{Telecom Infra Project}
\newacronym{tm}{TM}{Transparent Mode}
\newacronym{to}{TO}{Telco Operator}
\newacronym{toa}{ToA}{Time of Arrival}
\newacronym{tr}{TR}{Technical Report}
\newacronym{trp}{TRP}{Transmitter Receiver Pair}
\newacronym{ts}{TS}{Technical Specification}
\newacronym{tti}{TTI}{Transmission Time Interval}
\newacronym{ttt}{TTT}{Time-to-Trigger}
\newacronym{tx}{TX}{Transmitter}
\newacronym{uas}{UAS}{Unmanned Aerial System}
\newacronym{uav}{UAV}{Unmanned Aerial Vehicle}
\newacronym{udm}{UDM}{Unified Data Management}
\newacronym{udp}{UDP}{User Datagram Protocol}
\newacronym{udr}{UDR}{Unified Data Repository}
\newacronym{ue}{UE}{User Equipment}
\newacronym{uhd}{UHD}{\gls{usrp} Hardware Driver}
\newacronym{ul}{UL}{Uplink}
\newacronym{um}{UM}{Unacknowledged Mode}
\newacronym{uml}{UML}{Unified Modeling Language}
\newacronym{upa}{UPA}{Uniform Planar Array}
\newacronym{upf}{UPF}{User Plane Function}
\newacronym{urllc}{URLLC}{Ultra Reliable and Low Latency Communication}
\newacronym{usa}{U.S.}{United States}
\newacronym{usim}{USIM}{Universal Subscriber Identity Module}
\newacronym{usrp}{USRP}{Universal Software Radio Peripheral}
\newacronym{utc}{UTC}{Urban Traffic Control}
\newacronym{vim}{VIM}{Virtualization Infrastructure Manager}
\newacronym{vm}{VM}{Virtual Machine}
\newacronym{vnf}{VNF}{Virtual Network Function}
\newacronym{volte}{VoLTE}{Voice over \gls{lte}}
\newacronym{voltha}{VOLTHA}{Virtual OLT HArdware Abstraction}
\newacronym{vr}{VR}{Virtual Reality}
\newacronym{vran}{vRAN}{Virtualized \gls{ran}}
\newacronym{vss}{VSS}{Video Streaming Server}
\newacronym{wbf}{WBF}{Wired Bias Function}
\newacronym{wf}{WF}{Wired-first}
\newacronym{wi}{WI}{Wireless InSite}
\newacronym{wlan}{WLAN}{Wireless Local Area Network}
\newacronym{pnf}{PNF}{Physical Network Function}
\newacronym{drl}{DRL}{Deep Reinforcement Learning}
\newacronym{mtc}{MTC}{Machine-type Communications}
\newacronym{v2x}{V2X}{Vehicle-to-everything}
\newacronym{cast}{CaST}{Channel emulation scenario generator and Sounder Toolchain}
\newacronym{gui}{GUI}{Graphical User Interface}
\newacronym{ups}{UPS}{Uninterruptible Power Supply}
\newacronym{ota}{OTA}{Over-the-Air}
\newacronym{hitl}{HITL}{hardware-in-the-loop}
\newacronym{soc}{SoC}{System-on-Chip}
\newacronym{ecdf}{eCDF}{Empirical Cumulative Distribution Function}
\newacronym{cnn}{CNN}{Convolutional Neural Network}
\newacronym{nn}{NN}{Neural Network}
\newacronym{mpc}{MPC}{Multi Path Component}
\newacronym{otc}{OTC}{Over-the-Cable}
\usepackage{tikzpagenodes,etoolbox}
\usetikzlibrary{calc}
\usepackage[contents={}]{background}
\AddEverypageHook{%
\ifnumequal{\thepage}{1}{%
    \tikz[remember picture,overlay]{%
        \node[draw,
        minimum width=1.03\textwidth,
        text width=1.02\textwidth,
        font=\footnotesize
        ]
        at ($(current page header area) - (0,-6pt)$)
        {%
        This paper has been accepted for publication in the Proceedings of the IEEE International Workshop on Computer Aided Modeling and Design of Communication Links and Networks (CAMAD). This is the author's accepted version of the article. The final version published by IEEE is A. Saeizadeh, M. Tehrani-Moayyed, D. Villa, J. G. Beattie, Jr., I. C. Wong, P. Johari, E. W. Anderson, S. Basagni, T. Melodia, “AI-assisted Agile Propagation Modeling for Real-time Digital Twin Wireless Networks,” in Proceedings of the IEEE International Workshop on Computer Aided Modeling and Design of Communication Links and Networks (CAMAD), October 2024.
        };
    }%
}{}%end ifnumequal
}

\begin{document}

\title{AI-assisted Agile Propagation Modeling for Real-time Digital Twin Wireless Networks
\thanks{This work is supported in part by VIAVI Solutions, Inc., and by the National Telecommunications and Information Administration (NTIA)’s Public Wireless Supply Chain Innovation Fund (PWSCIF), Award No. 25-60-IF011.}
%
%\thanks{This work is partially supported by VIAVI Solutions, Inc., and by the National Telecommunications and Information Administration (NTIA)’s Public Wireless Supply Chain Innovation Fund (PWSCIF) under Award No. 25-60-IF011.}
}

\author{\IEEEauthorblockN{Ali Saeizadeh$^\dagger$, Miead Tehrani-Moayyed$^\dagger$, Davide Villa$^\dagger$, J.\ Gordon Beattie, Jr.$^*$, Ian C. Wong$^*$,\\Pedram Johari$^\dagger$,  Eric W. Anderson$^\dagger$, Stefano Basagni$^\dagger$, Tommaso Melodia$^\dagger$}

\IEEEauthorblockA{$^\dagger$Institute for the Wireless Internet of Things, Northeastern University, Boston, MA, U.S.A.\\
$^*$VIAVI Solutions, Inc.\\
E-mail: $^\dagger$\{saeizadeh.a, tehranimoayyed.m, villa.d, p.johari, er.anderson, s.basagni, melodia\}@northeastern.edu, \\$^*$\{gordon.beattiejr, ian.wong\}@viavisolutions.com
}}
\maketitle
\begin{abstract}
Accurate channel modeling in real-time faces remarkable challenge due to the complexities of traditional methods such as ray tracing and field measurements. 
%
%These methods are often resource-intensive and time-consuming. 
%
AI-based techniques have emerged to address these limitations, offering rapid, precise predictions of channel properties through ground truth data. 
This paper introduces an innovative approach to real-time, high-fidelity propagation modeling through advanced deep learning.
Our model integrates 3D geographical data and rough propagation estimates to generate precise path gain predictions. 
By positioning the transmitter centrally, we simplify the model and enhance its computational efficiency, making it amenable to larger scenarios.
Our approach achieves a normalized Root Mean Squared Error of less than~0.035~dB over a~37,210 square meter area, processing in just~46~ms on a GPU and~183~ms on a CPU. 
This performance significantly surpasses traditional high-fidelity ray tracing methods, which require approximately three orders of magnitude more time.
Additionally, the model's adaptability to real-world data highlights its potential to revolutionize wireless network design and optimization, through enabling real-time creation of adaptive digital twins of real-world wireless scenarios in dynamic environments.
%
%%% Previous abstract.
%Accurate channel modeling in real-time is a significant challenge due to the complexities of traditional methods such as ray tracing and field measurements. These methods are often resource-intensive and time-consuming. This paper introduces an innovative approach leveraging advanced \gls{dl} techniques to achieve real-time, high-fidelity propagation modeling. Our model utilizes elevation maps and rough propagation estimates to generate precise path gain predictions. By positioning the \gls{tx} centrally, we simplify the model and enhance its computational efficiency. Experimental results indicate that our model achieves a normalized \gls{rmse} 0.0268 dB over a 37,210 square meter area, processing in just 46 ms on a GPU and 183 ms on a CPU. This performance significantly surpasses traditional high-fidelity ray tracing methods, which require over \hl{387.6} seconds. This study underscores the potential of \gls{ai} in transforming wireless network design, providing substantial improvements in speed and accuracy for real-time propagation modeling.
\end{abstract}


\glsresetall

\section{Introduction}
\label{sec:intro}

\emph{Real-time} channel modeling is a cornerstone for modern and future wireless communication systems.
It allows for network design and optimization in a risk-free \gls{dt} environment to dynamically adapt to changing conditions of a real-world wireless network deployment. This enables better algorithmic designs and decision-making in a virtual environment, ultimately improving performance, efficiency, and user experience in real-world deployment. 
However, real-time modeling faces challenges due to the complexity of modern environments, including user mobility and varying conditions.
Traditional methods like field measurements, ray tracing, and stochastic models fall short. 
Field measurements are accurate but costly, ray tracing is detailed but computationally intensive (e.g., in large \gls{rf} scenarios with mobility~\cite{zhu2024digital}), and stochastic models are less resource-demanding but sacrifice accuracy~\cite{seybold2005introduction}.
Attempts to address these shortcomings have provided limited results. 
For instance, efforts to accelerate ray tracing, such as parallelizing algorithms and using GPUs, have not yet achieved real-time capabilities due to lengthy computation times. 
Consequently, real-time support for mobility models and optimization remains impractical~\cite{zhu2024digital}. 
Ray tracing also lacks flexibility: Scenario changes necessitate rerunning the process, complicating adaptation to dynamic environments with new transmitters or receivers in real-time. 
%
%Additionally, ray tracing cannot be refined using measurement data, limiting its effectiveness in dynamic and complex settings.

To address these limitations, advanced \gls{ai}-based techniques have been proposed. 
These methods leverage ground truth data from measurements or simulations to train data-driven models, enabling rapid and precise predictions of channel properties. 
By establishing a mapping function between the wireless environment and channel parameters, \gls{ai} tools facilitate proactive network design. 
They effectively tackle challenges such as resource allocation, user mobility analysis, localization, and radio propagation modeling. 
AI-based techniques offer greater flexibility, scalability, and reduced computational complexity, thus enabling real-time propagation modeling in complex urban environments~\cite{gupta2022machine, ratnam2020fadenet, levie2021radiounet, bakirtzis2022deepray, lee2023scalable, qiu2022pseudo, ozyegen2022empirical}.

Leveraging \gls{ai} for real-time modeling has multifold key motivations, including the need for advanced \gls{dt} technology~\cite{jones2020characterising, testolina2024boston} and enhanced telecommunication system design.
Real-time \glspl{dt} provide dynamic digital replicas of real-world network environments, enabling safe testing and evaluation of new configurations without impacting real-world performance~\cite{villa2024dt}.
They also facilitate data collection to train \gls{ai}/\gls{ml} models \cite{ alkhateeb2023real}. 
However, real-time \glspl{dt} for telecommunication networks face challenges due to increased deployment density, complex architectures, and high-frequency communications. 
Accurate real-time propagation modeling is crucial for enhancing the physical layer of these \glspl{dt}, ensuring they accurately reflect real-world conditions. 
Utilizing \gls{ai} for this purpose can significantly improve network optimization and performance management in dynamic telecommunication environments~\cite{khan2022digital}.
\gls{ai}-assisted propagation modeling enhances telecommunication system design by optimizing base station deployments, particularly RRHs or O-RUs, and visualizing signal propagation characteristics. 
When site conditions differ from the design, real-time modeling identifies discrepancies, enabling rapid adjustments to cell sites and sectors for optimal coverage and service.

Early efforts in \gls{ai}-based channel modeling utilized conventional ML techniques like Random Forest, KNN, and SVM to model channel path loss, with Random Forest showing the best performance~\cite{zhang2018air}. The introduction of \gls{cnn} advanced the field by leveraging spatial correlations in images to enhance model accuracy using propagation features like distance and building maps~\cite{imai2019radio}. Fully connected layers were used for regression, incorporating additional features such as frequency and antenna tilt. Further advancements included the use of satellite and aerial images as input channels, significantly improving prediction accuracy~\cite{hayashi2020study, thrane2020model}. Hybrid approaches combining rough estimates from physics-based models with \gls{nn} models also improved prediction performance~\cite{nguyen2022deep}.

Recent works have focused on incorporating more comprehensive input information and predicting channel parameter heat maps in one shot. \gls{cnn}-based Auto Encoders and U-Net networks have shown promising results for fast and accurate predictions, with U-Net providing better performance at the cost of increased computational demands~\cite{gupta2022machine, ratnam2020fadenet, levie2021radiounet, bakirtzis2022deepray, lee2023scalable, qiu2022pseudo, ozyegen2022empirical}. 
These advancements in \gls{ai}-based channel modeling highlight its potential to revolutionize wireless network design by enabling real-time, accurate propagation modeling across various scenarios and environments. However, despite their promising performance, questions remain regarding the generalizability of these models and their evaluation with real-world measurements. 

\begin{figure}[htb]
    \centering  \includegraphics[width=0.48\textwidth]{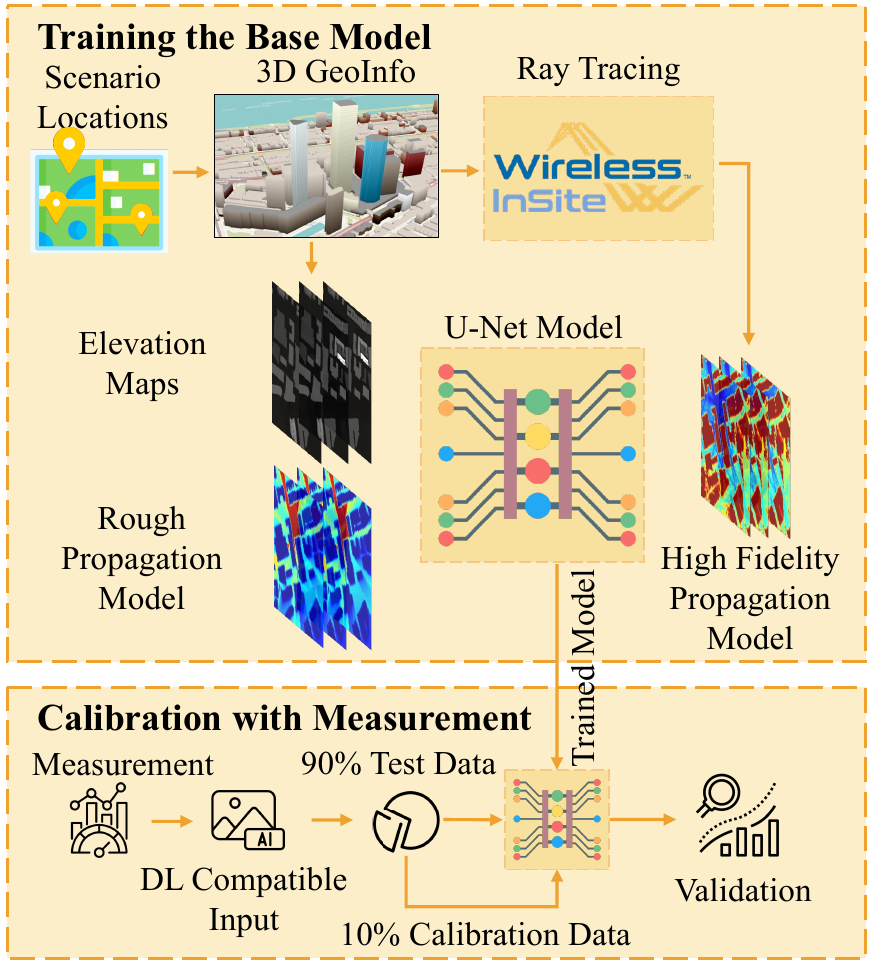}
    \caption{Training process, data calibration, and model validation.}
    \label{fig:diag}
    \vspace{-15pt}
\end{figure}

In this paper we address these concerns with the aim of enhancing the reliability of \gls{ai}-based approaches in diverse and dynamic wireless environments. 
To achieve this, we employ a U-Net structure inspired by Lee and Molish~\cite{lee2023scalable}, which allows the model to cover an entire area with a single inference while capturing spatial features effectively. 
We leverage two key inputs: an elevation map to accurately convey 3D geographical information to the model, and a rough estimation of the propagation model to maintain generalizability (Fig.~\ref{fig:diag}). 

This rough estimation is subsequently upsampled and refined into a high-fidelity propagation model. 
Unlike other approaches, we place the \gls{tx} at the center to reduce the model's complexity. 
This improves model performance compared to using an additional channel for the \gls{tx} location.
By implementing these changes, we 
%effectively 
address the generalizability issues of previous works, enabling real-time propagation modeling for any environment with available 3D geographical data. Our approach significantly improves both accuracy and computational efficiency. Specifically, we achieve a normalized Root Mean Squared Error (RMSE) of less than 0.035 dB over a 37,210 square meter area, with processing times of just 46 ms on a GPU. These results demonstrate the model's ability to rapidly provide high-fidelity propagation predictions, surpassing traditional ray tracing methods that require over $387.6$\;seconds. 
Additionally, refining the model with a small amount of measurement data shows an \gls{rmse} of $0.0113$, demonstrating its adaptability to real-world data.
%
%Additionally, refining the model with a small amount of measurement data shows an error of \hl{xx}, highlighting the its flexibility with measurement data. 
%
The remarkable performance and efficiency of AI-driven techniques underscore their potential to revolutionize wireless network design and optimization, enabling real-time adaptation to dynamic and complex telecommunication environments.

% summary of the paper
The paper is organized as follows. Section~\ref{sec:data} describes data collection for training the model. Section~\ref{sec:dl} details the architecture and components of the \gls{dl} model. Section~\ref{sec:results} presents performance results and an evaluation of the model. Finally, Section~\ref{sec:con} concludes the paper.

\section{Training the Model: Data Collection}
\label{sec:data}

To construct a comprehensive and precise dataset for predicting path gain (expressed in dB throughout this study), we utilized both ray tracing simulations and empirical measurements. Our objective is to forecast the Path Gain ($\mathrm{PG}$), as delineated in equation (\ref{eqn:pg_general}), by leveraging the received power at the receiver ($\text{P}_{\text{RX}}$) and the transmitted power ($\text{P}_{\text{TX}}$).
\begin{equation}
    \mathrm{PG}(t)
    = P_{\mathrm{RX}}(t) 
    - P_{\mathrm{TX}}(t),
    \label{eqn:pg_general}
\end{equation}
The \gls{cir} can be defined as:
\begin{equation}
h(t, \tau) = \sum_{i=1}^{N} \alpha_i(t) \delta\left(\tau - \tau_i(t)\right),
\label{eqn:cir}
\end{equation}
\noindent where \(\alpha_i(t)\) is the time-varying complex amplitude of the \(i\)-th path and \(\tau_i(t)\) is the time-varying delay of the \(i\)-th path.

To extract the received power from the channel, we use the \gls{cir} as shown in (\ref{eqn:p_rx})
\begin{equation}
\begin{aligned}
    P_{\text{RX}}(t) 
    &= 10 \log_{10} \left( \int_{-\infty}^{\infty}|h(t, \tau)|^2 d \tau \right) \\
    &= 10 \log_{10} \left( \sum_{i} \left|\alpha_i(t)\right|^2 \right).
\label{eqn:p_rx}
\end{aligned}
\end{equation}
In a stationary environment with a mobile receiver, $PG(t)$ and $\text{PG}(\boldsymbol{q}_{\mathrm{RX}})$, where $\boldsymbol{q}_{\mathrm{RX}}$ is the location of \gls{rx} can be considered equivalent since the position \(\boldsymbol{q}_{\mathrm{RX}}\) of the receiver varies with time \(t\). Thus, modeling \(\text{PG}(\boldsymbol{q}_{\mathrm{RX}})\) effectively captures the time-varying nature of the path gain \(\text{PG}(t)\) as the receiver moves through different locations \(\boldsymbol{q}_{\mathrm{RX}}\).

\subsection{Ray Tracing}
Urban environments pose challenges for conventional channel models, which often fail to accurately characterize channel properties. Ray tracing methods offer a potential solution for these complex scenarios. To build a comprehensive dataset, we employed Wireless InSite Ray Tracing software~\cite{remcom_wireless_insite} by RemCom to collect high-fidelity data. The ray tracing model is configured by one diffraction and four reflections. In the scenario described below, simulating one transmitter takes 1:25:12 using CPU or 0:03:16 using GPU on a machine with two Intel Xeon E5-2660 processors with 28 cores and one Nvidia Tesla K40c GPU with 2880 CUDA cores.
%MT: We are not providing our ray tracing parameter analysis in this paper. This contribution is described in the journal paper in details. 

%However, the computational demands of ray tracing are significant, particularly in large-scale, intricate environments with numerous transmitters and receivers—conditions necessary for training \gls{dl} models. Hardware platforms can struggle to handle this computational load efficiently. As a result, balancing ray tracing configurations with computational complexity is crucial to obtaining high-fidelity simulation results.
%In our experience with ray tracing tools, we estimate that incorporating at least one diffraction and eight reflections is necessary to accurately represent most \gls{nlos} locations, which results in \hl{x min of run-time on a yyy computer with zzz CPU or GPU and ddd memory.}

\begin{figure}
    \centering
    \includegraphics[width=0.47\textwidth]{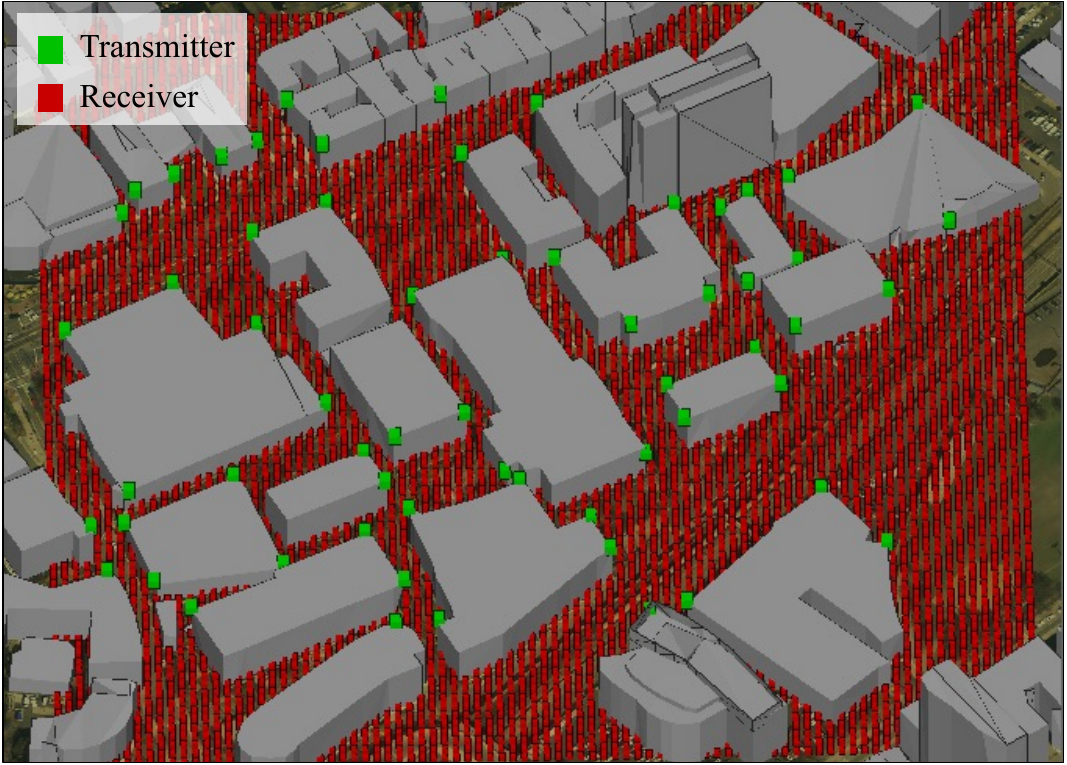}
    \caption{Northeastern University, Boston Campus, used for generating the main dataset consisting of 61 \gls{tx} and 7,569 \gls{rx}. Location: 42°20'22"N, 71°05'14"W.}
    \label{fig:neu}
    \vspace{-15pt}
\end{figure}

% Ray Tracing details 
In this study, we focus on the Northeastern University campus in Boston shown in Fig.~\ref{fig:neu}, as an urban use case scenario to train and test the model. We consider a grid of potential \gls{rx} locations, comprising 7,569 points (red points) spread over a $435 \times 435$ square meters area, and 61 \gls{tx} locations (green points) situated at the corners of building rooftops for potential \gls{bs}. Additionally, we examine Fenway Park (42°20'26"N 71°05'38"W), as a separate location with a lower density of buildings (and not seen in the \gls{dl} training phase) with another 16 \glspl{tx} to evaluate the model's generalization capabilities. To ensure accuracy, we import a precise 3D model of the area using data gathered by Boston Planning and Developing Agency \cite{BPDA_3D_Data_Maps}.

\begin{figure}[!b]
    \vspace{-15pt}
    %\centering
    \begin{subfigure}{0.23\textwidth}
        %\centering
        \includegraphics[width=.98\textwidth]{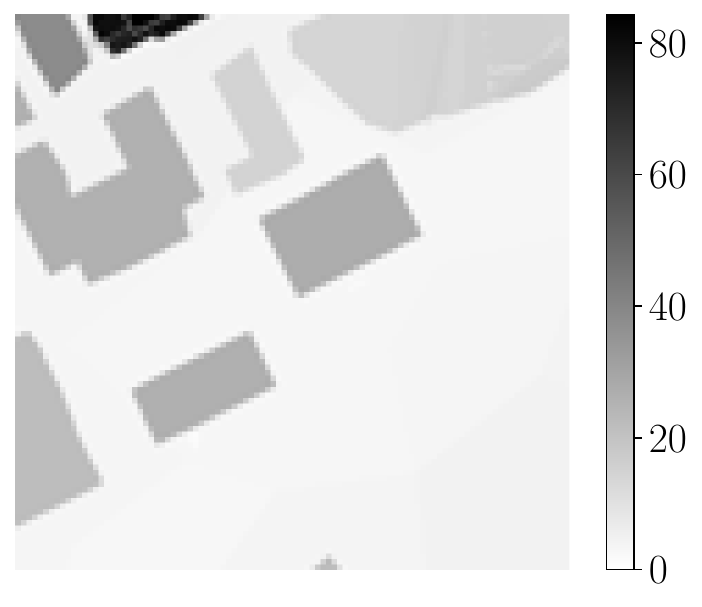}
        \caption{Elevation Map (M)}
        \label{fig:building_maps}
    \end{subfigure}
    \hfill
    \begin{subfigure}{0.24\textwidth}
        %\centering
        \includegraphics[width=.98\textwidth]{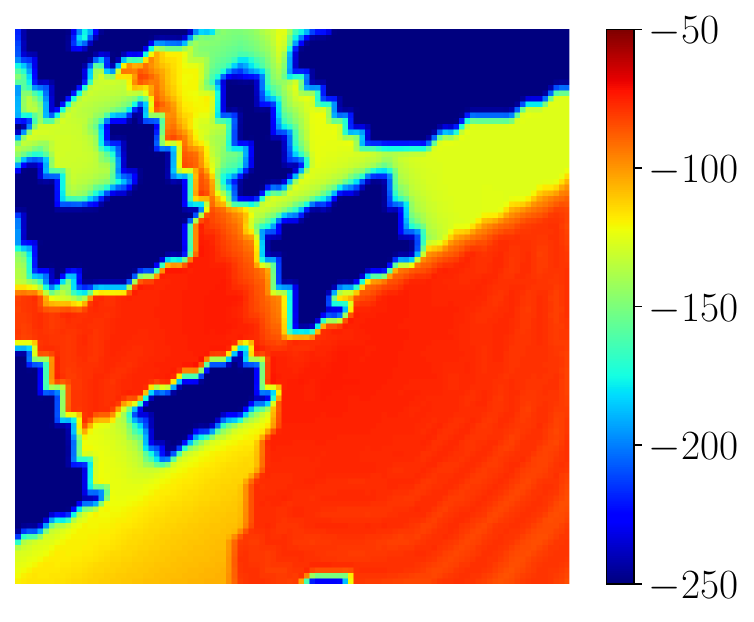}
        \caption{Propagation Estimate (dB)}
        \label{fig:rough}
    \end{subfigure}
    %\hfill
    \begin{subfigure}{0.24\textwidth}
        %\centering
        \includegraphics[width=.98\textwidth]{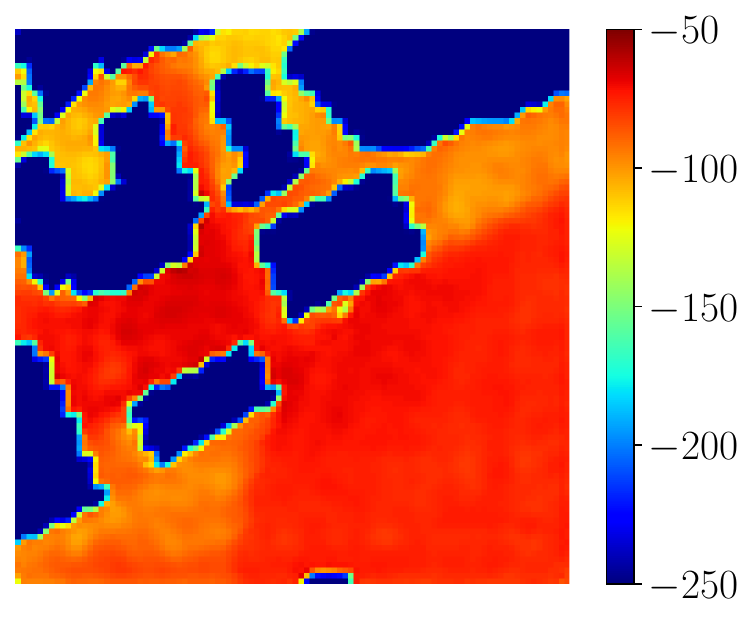}
        \caption{Model's Output (dB)}
        \label{fig:output}
    \end{subfigure}
    \hfill
    \begin{subfigure}{0.24\textwidth}
        %\centering
        \includegraphics[width=.98\textwidth]{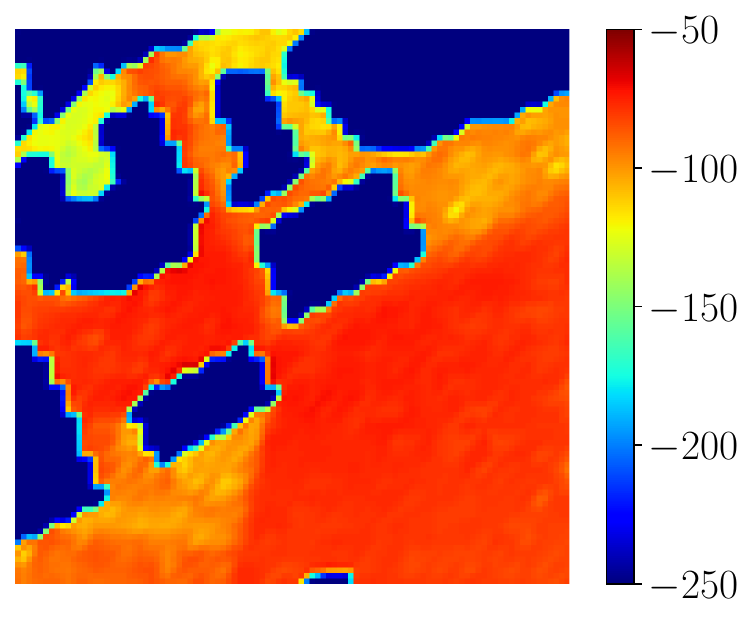}
        \caption{Ray Tracing (dB)}
        \label{fig:ground_truth}
    \end{subfigure}
    %\hfill
    \caption{A comparison of the model's output (Fig.~\ref{fig:output}) with the ground truth heatmap generated by Wireless InSite (Fig.~\ref{fig:ground_truth}), showing its superior performance compared to (Fig.~\ref{fig:rough}).}
    \label{fig:out}
\end{figure}

% Data Augmentation 
To prepare the input data for the model, we simulate ray tracing for all transmitter locations in the Northeastern University scenario, as shown in Fig.~\ref{fig:neu}, to obtain path gain heat maps. These heat maps serve as the ground truth for training the model, with Fig.~\ref{fig:out} presenting an example. We convert these maps into gray-scale single-channel images to reduce the data requirements and mitigate overfitting. Since \gls{dl} models require fixed input sizes, we crop the images to ensure uniform input sizes across different scenarios. Each image in our dataset measures $100\times100$ pixels, representing approximately a $1.929\times1.929$ meters area per pixel. Generating data for numerous scenarios is impractical, so we employ data augmentation techniques to create an augmented dataset from a limited synthetic dataset. Specifically, we use random rotations to incorporate the transmitter location into various input configurations, ensuring a substantial and diverse training dataset to enhance the model's robustness and accuracy.

\subsection{Measurement Campaign}
\begin{figure}[!t]
    \centering
    \begin{subfigure}{0.47\textwidth}
        \centering
        \includegraphics[width=\textwidth]{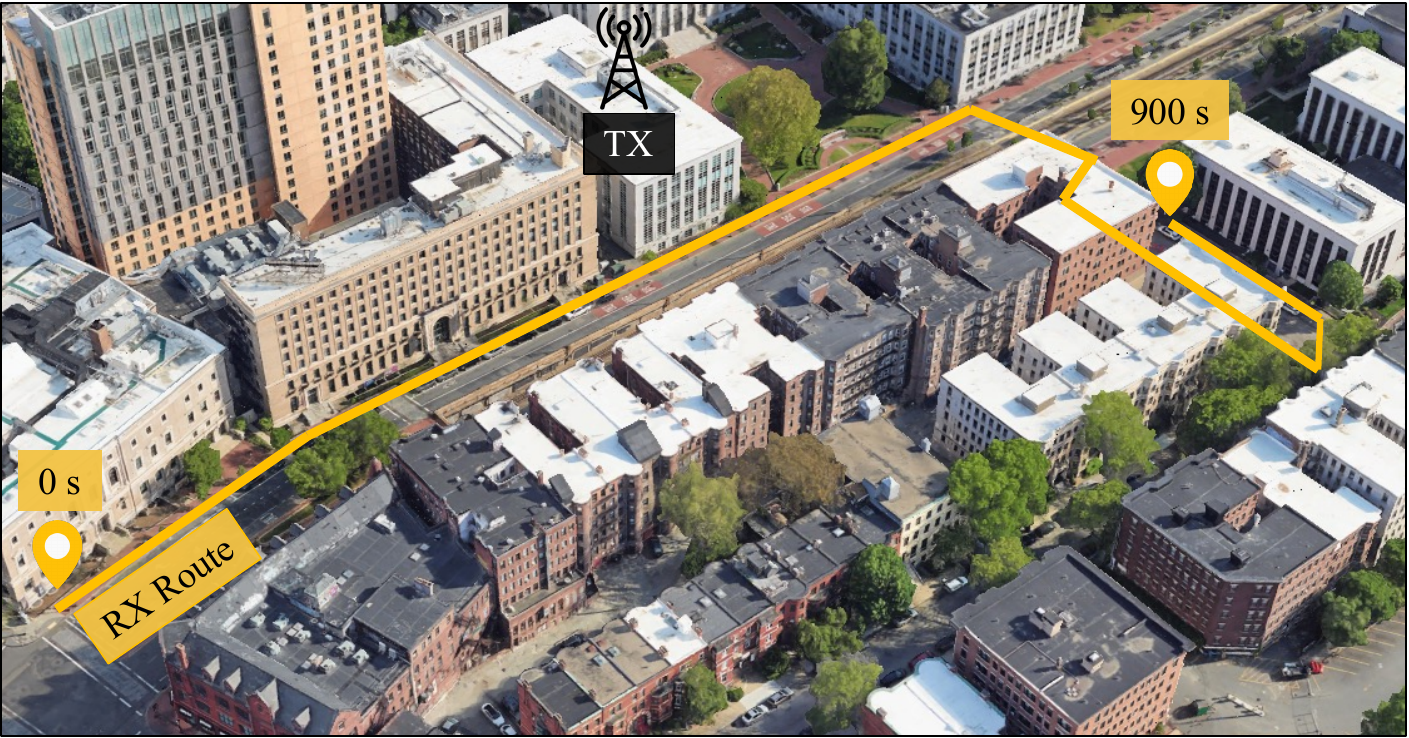}
        \caption{Path followed during the measurement campaign.}
        \label{fig:meas}
    \end{subfigure}
    \hfill
    \begin{subfigure}{0.47\textwidth}
        \centering
        \includegraphics[width=\textwidth]{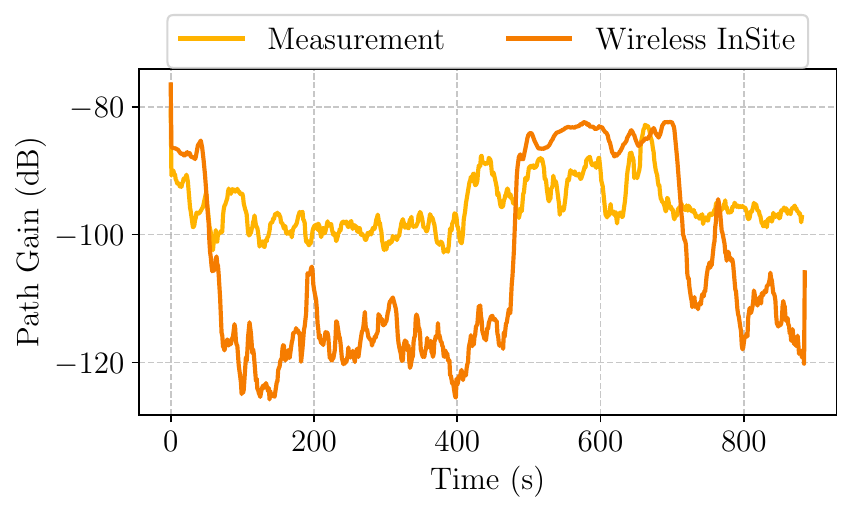}
        \caption{The path gain derived from the measurement campaign with moving average applied.}
        \label{fig:path_gain}
    \end{subfigure}
    \caption{Measurement campaign at Northeastern University Campus in collaboration with VIAVI to gather real-world data for model validation and refinement.}
    \label{fig:combined}
    \vspace{-15pt}
\end{figure}

% why ray tracing is not enough 
Although ray tracing is a powerful tool for modeling channels, real-world scenarios present unique variations that can significantly alter the channel characteristics (see Fig.~\ref{fig:path_gain}). Three main factors contribute to these variations: (1)~environmental dynamics, such as moving vehicles, constantly change the channel conditions; (2)~the precise shapes and materials of buildings, which 3D models may not accurately capture, can affect channel behavior, as materials are often not modeled with exact fidelity; (3)~interference from other users and background noise introduce additional environment noise into the channel, complicating accurate modeling.

% measurement campagin details 
To refine, calibrate, and validate the \gls{dl} model in a real-world scenario, we conducted a measurement campaign in collaboration with VIAVI Solutions around the Northeastern University campus, as depicted in Fig.~\ref{fig:meas}. The details of the measurements are provided in Table \ref{tab:meas-setup}.
\begin{table}[!h]
    \centering
    \caption{Measurement Setup and Equipment}
    \label{tab:meas-setup}
    \begin{tabular}{p{2.5cm}p{4.5cm}}
    \toprule
    \textbf{\textit{\gls{tx}}} & \\ 
    \midrule
    \gls{ru} & Ettus USRP X410 \\ 
    Amplifier & Minicircuits ZHL-1000-3W+ (38 dB) \\ 
    Antenna & Pasternack PE51OM1014 (6 dBi) \\ 
    Location & 42°20'25"N 71°05'16"W \\
    \toprule
    \textbf{\textit{\gls{rx}}} & \\ 
    \midrule
    \gls{ru} & Ranger provided by VIAVI Solutions \\
    Antenna & Waveform directional antenna (10 dBi) \\ 
    Location & Mobile (see Fig.~\ref{fig:meas}) \\
    \toprule
    \textbf{\textit{Measurement Details}} & \\
    \midrule
    Frequency & 910 MHz \\ 
    Bandwidth & 122.88 MHz \\ 
    Codeword & GLFSR-14 \\ 
    Synchronization & GPS clock for both \gls{tx} and \gls{rx} \\ 
    \bottomrule
    \end{tabular}
\end{table}

% Gordon
For channel sounding, the VIAVI Solutions Ranger, an \gls{rf} waveform generator and capture platform, was used. It supports two full-duplex channels with $200$\:MHz bandwidth up to $6$\:GHz and captures three and one-half hours of recordings. Analysis and waveform modifications are done using the Signal Workshop application, which can be run locally or remotely.

The recorded \gls{iq} samples by the Ranger were processed in two steps: (1) extracting \gls{cir} ($h(\tau,t)$) from the raw \gls{iq} samples ($R(\tau, t)$); and (2) calibration to determine the actual Path Gain ($PG(\mathbf{q}_{\mathrm{RX}})$). For the first step, we utilized a Galois Linear Feedback Shift Register 14 (GLFSR-14) codeword, modulated it using \gls{bpsk}, resampled it to match the \gls{rx} sampling frequency ($\mathbf{s}(t)$), and then correlated it with the received data. This process was repeated for each second of the data. Thus, \gls{cir} at the receiver $h[\tau,t=n]$ at the $n$-th second at location $\mathbf{q}_{\mathrm{RX}}$ using GPS logs is
\begin{equation}
    h(\tau,t) = R(\tau, t) \ast \mathbf{s}(t).
\end{equation}

To remove background noise from the channel, we identified the significant peaks in the channel for every length of the codeword:
\begin{equation}
    \mathbf{\alpha} = h(\tau, t) \cdot \mathbf{1}_{\{\text{peaks}(h(\tau, t))\}}.
    \label{eqn:peaks}
\end{equation}

The indicator function \(\mathbf{1}_{\{\text{peaks}(h(\tau, t))\}}\) is defined as
\begin{equation}
\mathbf{1}_{\{\text{peaks}(h(\tau, t))\}} =
\begin{cases} 
1 & \text{peaks of } h(\tau, t)  \\
0 & \text{otherwise.}
\end{cases}
\end{equation}
This function is $1$ at the peak (threshold of $+3 dB$ of the neighboring points) positions of \(h(\tau, t)\) and $0$ elsewhere, effectively isolating the peak values of the \gls{cir}. After extracting the sparse channel, we sum the values like in~(\ref{eqn:p_rx}).

To accurately determine the path gain, a conducted calibration test \gls{otc} was performed with all equipment in the loop except for the antennas. Instead of antennas, 60\,dB attenuators ($L_{\text{ATT}}$) were placed between \gls{rx} and \gls{tx} to prevent \glspl{adc} saturation, and the cable loss ($L_{\text{c}}$) was measured.
\begin{equation}
    P_{\text{\gls{rx}}}^{\text{\gls{otc}}}
    = P_{\text{\gls{tx}}} 
    + G_{\text{AMP}} 
    - L_{\text{c}} 
    - L_{\text{ATT}}
\end{equation}

After field measurements, the average received power from the calibration data ($P_{\text{\gls{rx}}}^{\text{\gls{otc}}}$) was subtracted from the received power in the actual measurement ($P_{\text{\gls{rx}}}^{\text{\gls{ota}}}$) to compensate for amplifiers ($G_{\text{AMP}}$), cable losses ($L_{\text{c}}$), and \gls{tx} power ($P_{\text{\gls{tx}}}$).
\begin{equation}
    P_{\text{\gls{rx}}}^{\text{\gls{ota}}}
    = P_{\text{\gls{tx}}}
    + G_{\text{AMP}}
    + G_{\text{ANT}}
    + \mathrm{PG}^{\text{\gls{ota}}}
\end{equation}

The nominal gain of the antennas ($G_{\text{ANT}}$) was also considered in $\mathrm{PG}^{\text{\gls{ota}}}$:
\begin{equation}
    \mathrm{PG}^{\text{\gls{ota}}}
    = P_{\text{\gls{rx}}}^{\text{\gls{ota}}}
    - P_{\text{\gls{rx}}}^{\text{\gls{otc}}}
    - L_{\text{c}}
    - L_{\text{ATT}}
    - G_{\text{ANT}}
\end{equation}

\section{DL-Based Propagation Model}\label{sec:dl}
\begin{figure}[t]
    \centering
    \includegraphics[width=0.48\textwidth]{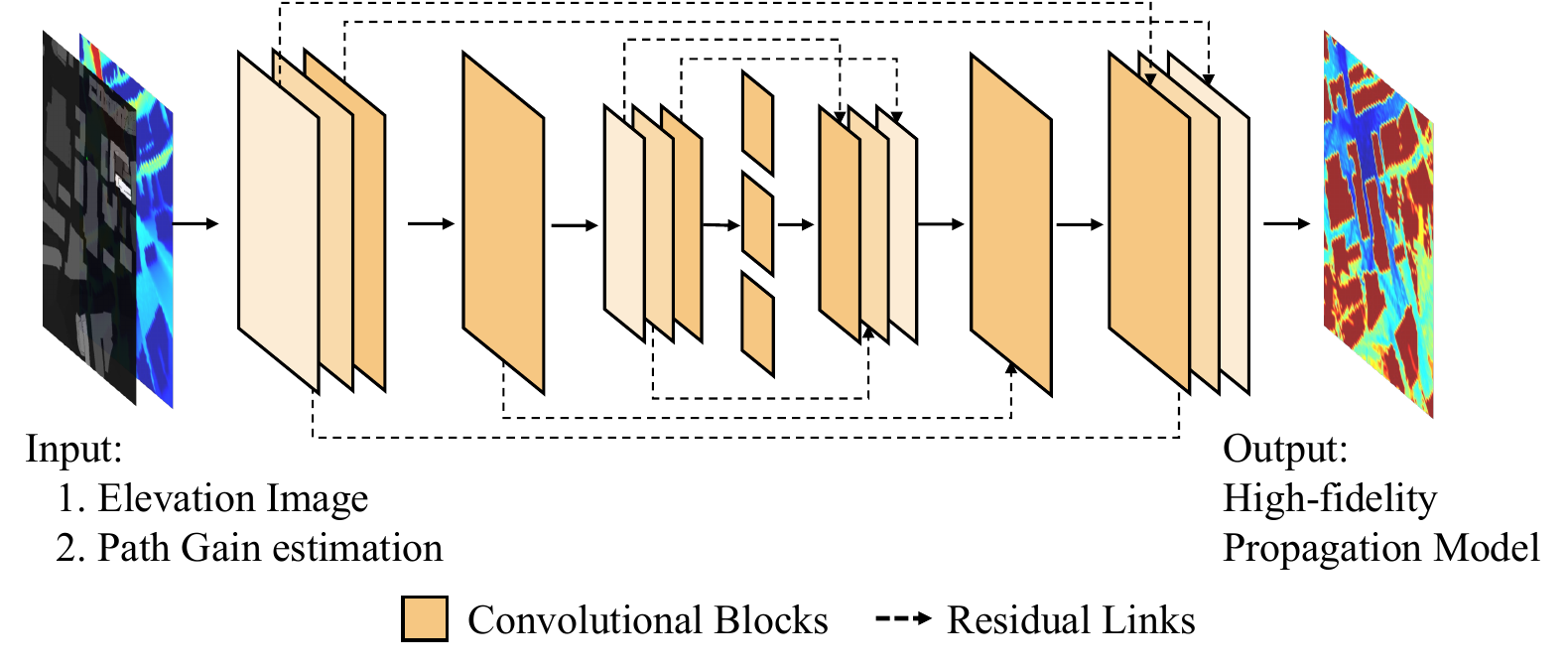}
    \caption{U-Net architecture adapted from PMNet~\cite{lee2023scalable}. We use elevation maps and propagation estimates as inputs to achieve an accurate propagation model.}
    \label{fig:dl_arch}
    \vspace{-15pt}
\end{figure}

% DL input output 
We propose a novel model shown in Fig.~\ref{fig:dl_arch} inspired by PMNet~\cite{lee2023scalable}, which originally takes two inputs: (1) a building map, indicating building locations with binary values (ones and zeros); and (2) a one-hot encoded \gls{tx} location. We have modified these inputs for improved performance. Instead of using a binary building map, we now input an elevation map ($\textbf{I}_{el}$) that shows building heights (Fig.~\ref{fig:building_maps}). Additionally, rather than a one-hot encoded \gls{tx} location, we use a rough estimation of the propagation model derived from real-time ray tracing by Wireless InSite ($\textbf{I}_{est}$), which runs in approximately $30$\:ms, as illustrated in Fig.~\ref{fig:rough}. The \gls{tx} location is always at the center of the image, so there is no need to include the \gls{tx} location in another channel. These modifications to the input enable our model to deliver high-fidelity propagation predictions in real time for different scenarios.

The \gls{dl} model is designed to predict the path gain ($PG_{dB}(\boldsymbol{q}_{\mathrm{RX}})$) given the input $\textbf{x}$ and the model parameters $\theta$ (weights and biases). However, the model's output is not limited to the path gain for a specific location, i.e., pixels, but rather it provides the path gain for an entire area, i.e., a heatmap. Mathematically, this can be represented as \(P(PG_{dB}(\boldsymbol{q}_{\mathrm{RX}}) | \textbf{x}, \mathbf{\theta})\), where $\textbf{x}$ is the concatenated input $[\textbf{I}_{el}, \textbf{I}_{est}]$, and $\theta$ encompasses all the parameters of the model. The objective is to learn the distribution of path gain conditioned on the input data and model parameters. This is achieved by optimizing the model parameters $\theta$ during training to minimize the prediction error.

As mentioned in Section~\ref{sec:data}, the main dataset consists of $61$~different scenarios (\gls{tx} locations), each with $100$~augmented scenarios, resulting in a total of $6,100$~images. To validate the robustness and ensure a fair comparison of the model, we randomly split the data into different ratios ten times. This approach helps to: (1) ensure there is no bias towards specific scenarios; and (2) evaluate the model's generalization across various test ratios. Additionally, the trained model is tested in a different type of environment at Fenway Park with 16 \glspl{tx} to further assess its performance. Besides ray tracing, measurement data is also used to validate the model. The entire process is illustrated in the diagram in Fig.~\ref{fig:diag}.

\section{Metrics and Results}\label{sec:results}

\begin{figure}[!b]
    \vspace{-15pt}
    \centering
    \begin{subfigure}{0.48\textwidth}
        \centering
        \includegraphics[width=\textwidth]{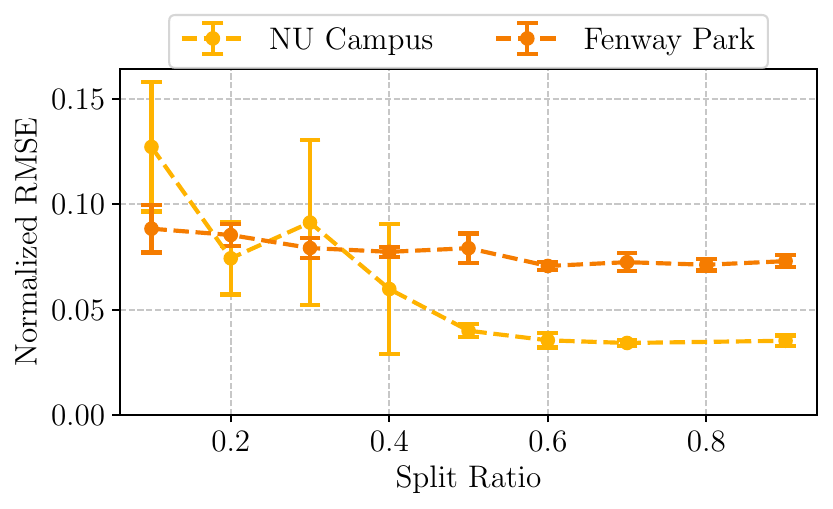}
        \caption{\gls{rmse} versus split ratio to evaluate the model's generalizability.}
        \label{fig:split}
    \end{subfigure}
    \hfill
    \begin{subfigure}{0.48\textwidth}
        \centering
        \includegraphics[width=\textwidth]{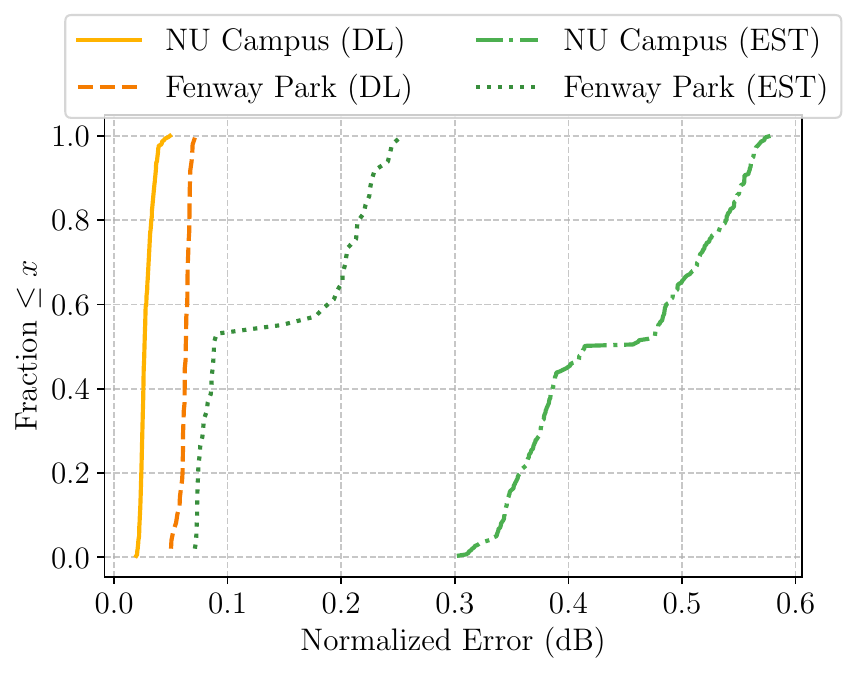}
        \caption{\gls{ecdf} of error for the \gls{dl} model compared to real-time propagation modeling from the Wireless InSite model (EST), illustrating the distribution of errors.}
        \label{fig:ecdf}
    \end{subfigure}
    \caption{Model performance comparison at Northeastern University and Fenway Park.}
    \label{fig:result}
\end{figure}

As described in Section~\ref{sec:dl}, the model is trained and tested with different split ratios using 10-fold cross-validation to find the optimal configuration. The results are shown in Fig.~\ref{fig:split}. As observed, the model stabilizes after a $0.6$~split ratio, indicating that no further training is required for effective inference. The error has been normalized across all plots, with the path gain range spanning from $-50$ to $-250$\:dB.

Also, the \gls{ecdf} of error for the best-performing model across 10-fold cross-validation has been plotted using two different test datasets (unseen by the model): (1) the same area at Northeastern; and (2) a different area at Fenway Park (42°20'26"N, 71°05'38"W), where the environment is more open compared to the dense structure of the Northeastern Campus. 

Results in Fig.~\ref{fig:ecdf} show that the model outperform traditional propagation estimation. Specifically, the median error for \gls{dl} at Northeastern Campus is $0.0268$, and at Fenway Park it is $0.0484$. However, for the traditional model, the median error at Northeastern Campus is $0.4146$, and at Fenway Park it is $0.0778$. These results highlight the model's robustness and adaptability to different environments, maintaining superior performance in both familiar and new scenarios.

\begin{figure}[!t]
    \centering
    \includegraphics[width=0.48\textwidth]{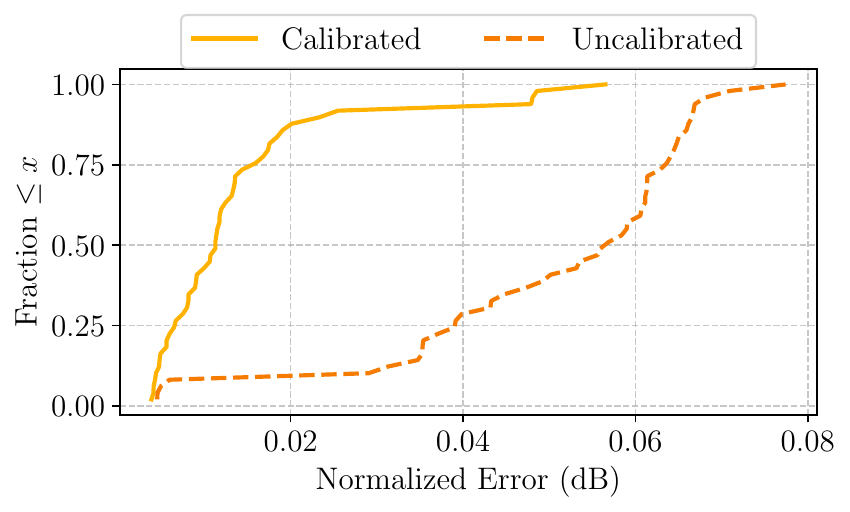}
    \caption{\gls{ecdf} of error w/ and w/o calibration using the measurement data (10\% of the dataset).}
    \label{fig:ecdf-meas}
    \vspace{-15pt}
\end{figure}

In addition to ray tracing, we evaluate the model's performance after adding measurement data in the training phase (Fig.~\ref{fig:ecdf-meas}). For this, we use 90\% of the points for testing. Initially, the error is relatively high (median of $0.0569$\:dB), but after refining the model with a small amount of measurement data, the performance improves significantly, reducing the median error to $0.0113$\:dB. This demonstrates that the \gls{dl} model can be calibrated with measurement data, unlike common ray tracing software.

Besides the low overall error of the propagation model, using \gls{dl} models enables near real-time accurate propagation modeling. From our test benches, real-time ray tracing by Wireless Insite takes approximately $30$\:ms to run, while high-fidelity ray tracing takes over $387.6$\:s to complete. However, %considering the rough estimation,
our model requires only $46$\:ms on a GPU and $183$\:ms on a CPU to run with minimal error.

\section{Conclusions}\label{sec:con}
We introduce a novel real-time path gain estimator leveraging advanced deep learning techniques. 
Our approach integrates elevation maps and rough propagation model estimations for highly accurate propagation modeling. 
Placing the transmitter at the center simplifies the model while enhancing performance, with precise path gain predictions in near real-time.
Our experimental findings demonstrate that the proposed model achieves a normalized RMSE of less than $0.035$\:dB across a $37,210$\:square meter area, executing in just $46$\:ms on a GPU and $183$\:ms on a CPU. 
This represents a significant advancement over traditional high-fidelity ray tracing methods, which typically require over $387.6$\:seconds on a GPU to complete.
Our research underscores the potential of AI-enhanced techniques to transform wireless network design. 
It lays a foundation for real-time digital twins, promising  efficient deployment and maintenance of future wireless infrastructure.

%In this paper, we introduced a novel real-time accurate path gain estimator leveraging advanced \gls{dl} techniques. Our approach utilizes elevation maps and rough propagation model estimations to achieve high-fidelity propagation modeling. By placing the \gls{tx} at the center, we reduce model complexity and enhance performance, demonstrating the capability to provide precise path gain predictions in near real-time. Our experimental results indicate that the proposed model achieves an normalized \gls{rmse} of less than 0.035 dB over a 37,210 square meter area in just 46 ms on a GPU and 183 ms on a CPU. This is a significant improvement compared to traditional high-fidelity ray tracing methods, which take over 387.6 seconds to run. Overall, our work highlights the potential of \gls{ai}-enhanced techniques to revolutionize wireless network design, providing a foundation for the development of real-time \glspl{dt} and enabling more efficient deployment and maintenance of wireless infrastructure.

%%% SB: After the paper is accepted.
%\section*{Acknowledgment}

\bibliographystyle{IEEEtran}
\bibliography{biblio}

\end{document}